# Applying the Chinese Wall Reverse Engineering Technique to Large Language Model Code Editing

**Manatsawin Hanmongkolchai**
chinesewall@whs.in.th

### Abstract

Large language models for code (Code LLM) are increasingly utilized in programming environments. Despite their utility, the training datasets for top LLM remain undisclosed, raising concerns about potential copyright violations. Some models, such as Pleias and Comma put emphasis on data curation and licenses, however, with limited training data these models are not competitive and only serve as proof of concepts.

To improve the utility of these models, we propose an application of the "Chinese Wall" technique, inspired by the reverse engineering technique of the same name – a high quality model is used to generate detailed instructions for a weaker model. By doing so, a weaker but ethically aligned model may be used to perform complicated tasks that, otherwise, can only be completed by more powerful models.

In our evaluation, we've found that this technique improves Comma v0.1 1T's performance in CanItEdit benchmark by over 66%, and Starcoder2 Instruct by roughly 20% compared to when running the same model on the benchmark alone. The practical application of this technique today, however, may be limited due to the lack of models trained on public domain content without copyright restrictions.

## 1 Introduction

Large language models for Code (Code LLM) are becoming more used to write programming code, with commercial code editors such as Cursor and GitHub Copilot using various commercially available models such as OpenAI's GPT-4, o3 or Google Gemini 2.5 Pro to perform the various tasks related to programming. However, the training data source for those models are not publicly available and it is commonly believed that the training data set does contain open source code with various licenses. Experiments such as Armin Ronacher [1]'s 2021 attempt with GitHub Copilot demonstrates that the tool is able to reproduce Quake's fast inverse square root algorithm in full, including comments. Quake's source code is licensed under the GNU General Public License v2.0 (GPLv2) which place conditions on the use of the source code. By outputting copyrighted code without attribution, the end user may unknowingly violate the source code license[1]. One method used by GitHub Copilot to prevent this issue is to perform a code search on the generated output to see if it matches any known public source code,

---
[1]This work does not provide legal advice, and do not claims that any legal opinion provided are correct

and block the model's output accordingly. This technique might not completely block partial matches and does not work for open source development as reproduction of the original code is expected.

Some models address this issue by using curated datasets with appropriate licensed contents. For example, the Stack v2 dataset [2] and Starcoder2 model limits data to permissively licensed sources and contents with unknown license. The Common Pile dataset [3] and the accompanied Comma model improves on this by limiting the dataset to permissive licensed contents only. Most permissive licenses only have attribution as its primary sole licensing condition and may be easier to comply with than the GPLv2 license. Ideally, models that are trained on public domain contents may be the best in terms of legal compliance as they have no restrictions or requirements, but to our knowledge no such text generation models exist today with reasonable quality. Additionally, the Common Pile authors wrote that *"license laundering"* – redistribution of copyrighted works with incorrect license – is commonly found on the internet, which may contaminate models that otherwise contain permissively-licensed code only.

In this work, we propose a "Chinese Wall" technique inspired by the similarly named reverse engineering technique [4]. The technique traditionally involves two teams of engineers. Initially, the first team performs reverse engineering on proprietary artifacts and writes detailed specifications of them. The second team, with no prior knowledge of the proprietary system, then studies the documents produced by the first team, and produces working software according to specifications. The resulting product, even if it's a perfect clone of the original systems' functionalities, may be considered legally distinct as has been claimed by Phoenix Technologies' clone of the IBM PC BIOS [4] among others.

We apply this technique to Code LLM by using high quality proprietary models to produce detailed instructions as inline comments, then using weaker models to follow the commented instructions in hope that it improves the weaker model's performance. We do not claim, however, that the process will always completely satisfy every legal requirement that allowed the original reverse engineering technique to produce legally distinct code.

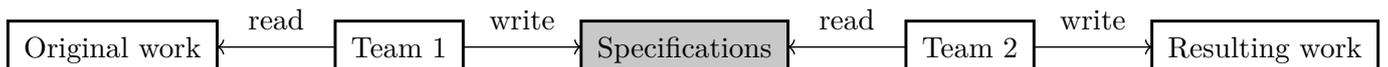

Figure 1: Traditional "Chinese Wall" reverse engineering flow

## 2 Related Work

A similar technique has been used by Aider's Architect mode [5] to achieve different goals. The Architect mode uses pair programming technique by using a strong reasoning model to act as the "architect". A separate "editor" model receives the architect's instructions and performs the actual edits accordingly. The intention is to utilize a strong reasoning model that is able to understand the codebase well, but perform poorly at generating *machine-parsable* editing instructions alongside another model that is adept at formatting the edit instructions. In this mode, the architect model is allowed to output actual code. Aider reports in the initial release that using this technique with o1-preview as the architect, and Deepseek as the editor has set a new bar in their code editing benchmark.

# 3 Models Used

The following models are used. Note that the runtime in use may inject its own system prompts, which may not be present in sample outputs.

| Model | Quantization | Version | Runtime |
| --- | --- | --- | --- |
| Comma v0.1 1T | BitsAndBytes 8B | 30dd729 | vLLM 0.9.2.dev312+g7b1895e6c |
| starcoder2:instruct | GGUF Q4_0 | 432973cfbc4c | Ollama 0.9.2 |
| phi4 | GGUF Q4_K_M | ac896e5b8b34 | Ollama 0.9.2 |
| google/gemini-2.5-pro | — | N/A | Google via Requesty.ai |

Table 1: Models used

## 3.1 Open Weight or Open Source Models

Models are chosen from their claims that the authors have paid attention to training data's license, but may not consist solely of permissively licensed data.

- **Comma** [3] is a 7 billion parameters language model. It was trained on The Common Pile database [3] which includes the permissively licensed part of The Stack v2 database, among other sources. Compared to Starcoder2 and The Stack v2, the creation of The Common Pile database is more focused on the licensing of the data used. This model is a base model and is not tuned to specific uses, such as instruction-following or code completion.

  The Comma model we use is the version trained from 1 trillion tokens. There's also a version trained from 2 trillion tokens, which is mostly a duplication of the base 1 trillion tokens. The 1T version is then quantized using 8-bit BitsAndBytes [6] to reduce memory usage to fit into the GPU we have on hand. This quantized version is published on our Hugging Face repository [7].

- **Starcoder2 Instruct** [8] is a 16 billion parameters language model trained to follow human written instructions. The instruction-tuning model was trained from the original Starcoder2 model [2] using the SelfCodeAlign pipeline proposed by Yuxiang Wei *et al.* [8]. The pipeline first generates coding concepts from seed functions in The Stack [9], a permissively licensed code dataset. Then, the base model is asked to generate new coding instructions. The model then has to follow the self-generated instructions to generate both responses, and test cases. Only successful cases are used to perform instruction tuning. Using this method means that the resulting model does not get contaminated from external data. The original Starcoder2 model was trained from The Stack v2 database [2], which include text and code from various sources that includes both permissively-licensed sources, and sources without explicit license. Authors of incorporated works can verify the inclusion of their works and opt out of The Stack database by using the "Am I in the Stack" tool, but the process is not instant nor reflected to existing versions of the model.

- **Phi-4** [10] is a 14 billion parameters language model from Microsoft Research. The model uses undisclosed training dataset, and is said to be trained on "a blend of synthetic datasets, data from filtered public domain websites, and acquired academic books and Q&A datasets". The copyright information for any data used to train the model is not available, including the list of sources and their license. While this model does not align to our goals, it is chosen as the best open weight model that has *some* care put into honoring some training data source's licenses.

## 3.2 Commercial Model

- **Gemini 2.5** [11] is a family of multimodal models by Google DeepMind. The family has three members: Flash Lite, Flash, and Pro. The Flash and Pro models support thinking ability and a very large context window. They're additionally trained to focus on programming tasks, such as IDE functionalities, coding agents, and end-to-end use for web & mobile application development. In Aider's Polyglot leaderboard [12] and LM Arena WebDev Arena leaderboard [13], Gemini 2.5 Pro is the highest rated model at the time of writing.

## 3.3 Other Models of Interests not in Use

There are other models of interests, but are not tested as part of this work due to various reasons:

- **Starchat2** is a 16 billion parameters language model trained by Lewis Tunstall *et al.* [14]. This model uses the OpenOrca dataset and UltraFeedback dataset to perform additional instruction tuning on Starcoder2. Both dataset are collections of completions from proprietary models such as GPT-3.5 and GPT-4. While OpenAI's Terms of Service [15] states that the user owns the output, the legality of OpenAI's output is being challenged [16], [17], [18] it is unclear whether such statement is valid.
- **Octocoder** is a 15.5 billion parameters language model trained by Niklas Muennighoff *et al.* [19] from the first Starcoder model [20]. The training dataset comes from the CommitPackFT database, a database of filtered permissively-licensed GitHub commit messages that resembles instructions. This database is also part of The Common Pile database that is used in training Comma.
- **Pleias 3B (Preview)** [21] is a model in early development by Pleias on the Common Corpus dataset. The model series is claimed to be the first models to be trained exclusively on permissively licensed data to follow the requirements of EU AI Act's Code of Conduct. However, Common Corpus includes Wikipedia contents which are mostly licensed in "copyleft"-style licenses — any downstream usage must retain similar copyright terms. This limits the use of the data in proprietary environments. From our brief testing, we were unable to utilize this model to reliably output code.

# 4 Evaluation

## 4.1 The CanItEdit Benchmark

Federico Cassano *et al.* [22] proposed the CanItEdit benchmark, consisting of 105 hand-crafted instructional Python code editing problems. Each problem has a "before" code, an "after" code (expected result), a test suite and two prompts variants – "descriptive", and "lazy". The evaluator software is published on the project's GitHub.

**Baseline** We modified the project's benchmark to add support for Ollama in addition to vLLM, as this is the runtime commonly used by end users in code editors. The original prompts are used with adaptations to fit Ollama's chat format. The hyperparameters used is the same as the original work: 2048 maximum new tokens, temperature 0.2, top-$p$ sampling cutoff of 0.95. For each problem, 20 completions are sampled.

During the test we've found that Gemini 2.5 Pro API will not return any model response if the maximum token is hit, which it has done so in 80 test cases. We retried those cases with 100,000 maximum tokens instead.

**Implementation**  Once the baseline result is established, we implement our Chinese Wall technique using Google Gemini 2.5 Pro in tandem with open weight models using one-shot prompts for both models. First, we pass the original task description and initial source code to Gemini, asking it to annotate the code with instructional comments. The Gemini hyperparameters are: 100,000 maximum tokens, temperature 0.2, top-$p$ sampling cutoff of 0.95, default thinking. We then pass the Gemini-annotated code into the editor model, along with the problem description. Other than the code, other outputs from the stronger model are not visible to the editor model. For each problem, we will only create one single annotated code from Gemini that is fed into all editor models. Each editor model will generate 20 sampled completions per problem. To reduce the variability of the test, and to save on the cost of doing the evaluation, the Gemini-outputted annotated source code is reused between all editor models.

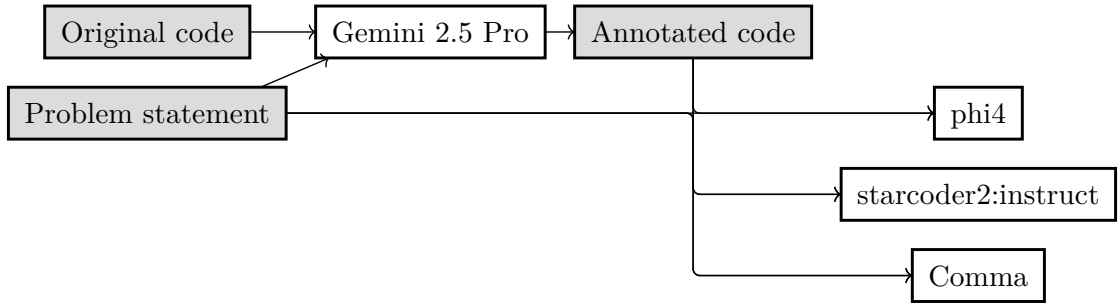

Figure 2: Implementation data flow

### 4.1.1 Results

| Model | Descriptive | | | Lazy | | |
|---|---|---|---|---|---|---|
| | pass@1 | pass@20 | ExcessCode | pass@1 | pass@20 | ExcessCode |
| **Gemini 2.5 Pro + phi4** | **68.67** | **72.38** | **0.14 ± 0.06** | **55.33** | **60.00** | **0.28 ± 0.19** |
| Gemini 2.5 Pro | 57.66 | 94.29 | 0.31 ± 0.09 | 53.60 | 86.67 | 0.41 ± 0.15 |
| phi4 | 54.71 | 73.33 | 0.47 ± 0.17 | 49.62 | 66.67 | 0.42 ± 0.13 |
| **Gemini 2.5 Pro + starcoder2:instruct** | **42.05** | **71.43** | **0.19 ± 0.08** | **34.67** | **55.24** | **0.30 ± 0.19** |
| starcoder2:instruct | 35.10 | 60.00 | 0.37 ± 0.13 | 26.00 | 57.14 | 0.21 ± 0.13 |
| **Gemini 2.5 Pro + Comma** | **21.24** | **33.33** | **0.23 ± 0.16** | **17.14** | **26.67** | **0.11 ± 0.10** |
| Comma v0.1 1T | 9.14 | 20.00 | 0.86 ± 0.60 | 7.67 | 10.48 | 0.21 ± 0.20 |

Table 2: CANITEDIT evaluation result

**Evaluation metrics**  The *pass@k* metric is the likelihood that at least one successful edit (with all tests passed) was made from *k* attempts. In other words, *pass@1* is the percent of passing attempts in all 2,100 samples, and *pass@20* is the number of passing problems where at least 1 edit attempt passed all test cases. The *ExcessCode* metric measures the number of unnecessary code changes as indicated by the number of changed lines not covered by the test suite. The original work [22, p 7-8] has additional details of how the metrics are calculated.

From the result, we can see that with the aid of Gemini 2.5 Pro, Comma v0.1 1T were able to complete the task more reliably, with *pass@1* result improving by 120%. At *pass@20* its ability were improved by 66%. Starcoder2 Instruct's ability were improved by roughly 20%.

As for Phi-4, the *pass@1* result improved by 25%, exceeding that of Gemini 2.5 Pro. However, its *pass@20* result becomes slightly reduced, which shows that this technique may not always improve the capabilities of a stronger model.

## 5 Conclusion

In this work, we proposed the "Chinese Wall" technique for enhancing weaker language models for code (Code LLM) at inference time by using a more powerful LLM to generate descriptions of the task. While we're unable to go into details about how legally distinct the resulting work is compared to when using just one language model, the technique shows promises that it improves the ability of a weaker language model but is still behind the more powerful model itself.

The application of this technique in the real world today, however, is limited. As Comma and Pleias are the only model known to us today that is trained solely from permissive data sources, and there's no models that were trained solely on public domain sources, it is not possible to use this technique with those models to ensure that the outputting LLM have no knowledge of copyrighted materials with license restrictions. We hope that it may be useful when such model exists in some capacity.

## 6 Acknowledgements

The author would like to thank the Tildes community for bringing the Comma model to our attention, which enables this idea to become a possibility. I'd also like to thank Assoc. Prof. Dr. Jittat Fakcharoenphol from Kasetsart University, my alma mater, for his assistant in publishing my first paper.

# APPENDIX A Quake's "Fast Inverse Square Root" test

In our initial test to prove the viability of this technique, we tried a modified version of Armin Ronacher [1]'s 2021 Copilot Fast Inverse Square Root test for conversational models. This algorithm is famously used in Quake III Arena [23, L552-572] to quickly estimates $\frac{1}{\sqrt{x}}$, and is well-known for its "magic" constant `0x5F3759DF`.

First, we used GPT 4o mini [24] on Kagi Assistant (with the web function disabled) to produce the annotated code. The prompt used is modified from CanItEdit's prompt.

**User**

```
You are CTeacherGPT. You will be provided the original code snippet and
an instruction that specifies the changes to complete the task. You must
not complete the task. Instead, you will add comments to the code, and
NEVER any code. Each comments is a step-by-step, detailed instructions so
that a novice programmer could follow to complete the task without
reading the original instruction. Your comments must be placed at the
spot where you would normally make such edit. You may make multiple
comments. Your comment must starts with `EDIT:`. Only produce the
commented code, do not include any additional prose.

## Instruction
Implement a fast inverse square root function in C
```

**Assistant**

```c
float fast_inverse_sqrt(float number) {
  // EDIT: Start by defining a variable to hold the result of the bit manipulation.
  float result;

  // EDIT: Use a union or a pointer to manipulate the bits of the float number.
  // This will allow us to access the float's binary representation as an integer.

  // EDIT: Perform the bit manipulation to approximate the inverse square root.
  // This typically involves using a magic number and shifting the bits.

  // EDIT: Use the result of the bit manipulation to refine the approximation.
  // Implement one or two iterations of Newton's method for better accuracy.

  // EDIT: Return the final result, which is the fast inverse square root of the input number.
  return result;
}
```

**Using the result on `starcoder2:instruct`**

We fed the GPT 4o mini's output into `starcoder2:instruct` running in Ollama. The model does not receive the user's instructions other than as put into the code by GPT 4o mini.

| **User** |
| Update this code by following the instruction in all comments that starts with `EDIT: `. Then remove the EDIT comments. Only produce the code, do not include any additional prose.<br><br>## Code<br>float fast_inverse_sqrt(float number) {<br><br>*(Code from previous response omitted for brevity)* |
| **Assistant** |
| *(Response shortened for brevity)*<br><br>```c<br>float inv_sqrt(float number) {<br>  float result;<br>  union {<br>    float f;<br>    uint32_t i;<br>  } u;<br>  u.f = number;<br>  u.i = 0x5F3759DF - (u.i >> 1);<br>  result = u.f;<br>  result = 0.5 * (result * (3.0 - number * result * result));<br><br>  return result;<br>}<br>``` |

Compared to the original code used in Quake, the Starcoder2's generated algorithm use C `union` instead of pointer casting, use `uint32_t` instead of `long`.

**Comparison with unaided `starcoder2:instruct`**

| **User** |
| Write fast inverse square root algorithm in C |
| **Assistant** |
| *(Response shortened for brevity)*<br><br>```c<br>float fast_inverse_sqrt(float x) {<br>  float xhalf = 0.5f * x;<br>  int i = *(int*)&x;<br>  i = 0x5f3759df - (i >> 1);<br>  x = *(float*)&i;<br>  x = x * (1.5f - xhalf * x * x);<br><br>  return x;<br>}<br>``` |

This generated code is more similar to the Quake's code.

**Using the result on Comma v0.1 1T (8-bit BitsAndBytes quantization)**

We fed the same result into Comma, with a different prompt as it is not an instruction-tuned model.

| User |
|---|
| ## Code Before:<br>def add(a, b):<br>  return a + b<br>## Instruction:<br>Add a "sub" function that subtracts two numbers. Also write docstrings for both functions and change a,b to x,y.<br>## Code After:<br>def add(x, y):<br>  """Adds two numbers."""<br>  return x + y<br><br>def sub(x, y):<br>  """Subtracts two numbers."""<br>  return x - y<br>## Code Before:<br>float fast_inverse_sqrt(float number) {<br><br>*(Code omitted for brevity)*<br><br>}<br>## Instruction:<br>Implement a fast inverse square root function in C<br>## Code After: |
| **Assistant** |
| ```c
float fast_inverse_sqrt(float number) {
    float result = 0.0f;
    // Use bitwise operations to extract the exponent and mantissa
    result = number & 0x7fffffff;
    result = (number >> 23) - (result >> 15);
    result = (number & 0x3f800000) >> 27;
    return result;
}
```<br>*(Rest omitted as stop tokens are not set)* |

The algorithm doesn't appear to be a valid one.

**Comparison with unaided Comma**

| User |
|---|
| ## The fast inverse square root algorithm in C<br><br>float fast_inverse_sqrt(float x) { |
| **Assistant** |
| ```c
        float r = 1.0f / x;
        float t = 1.0f / r;
        float r2 = r * r;
        float t2 = t * t;
        float r3 = r2 * r;
        float t3 = t2 * t;
        float r4 = r3 * r;
        float t4 = t3 * t;
``` |

> *(Rest of the response omitted)*

It seems that the model does not recognize the algorithm and produces a useless chain of variables. When asked for normal "inverse square root algorithm" without "fast", it returns `float inverse_sqrt(float x) { return x * 0.5f; }`, which is still incorrect.

# APPENDIX B Sample CANITEDIT outputs

## B.1 Test case 10 — CSV Parser

**Prompt Type** Descriptive

**Instructions** Add a function called `header` which returns the first row of a csv file as a list of strings, where every element in the list is a column in the row.

**Sample output of Gemini 2.5 Pro as annotator**

```python
class CSVParser:
    def __init__(self, csv: str):
        self.csv = csv

    def contents(self) -> list[list[str]]:
        lines = self.csv.split("\n")
        output = []
        for line in lines:
            output.append(line.split(","))
        return output

    # EDIT: Define a new function here called `header`.
    # It should take `self` as its only argument.
    # It should also have a return type hint, indicating that it returns a list of
strings (`list[str]`).

        # EDIT: The goal of this function is to return the first row of the CSV.
        # You can get all the rows by calling the `self.contents()` method, which
returns a list of lists.

        # EDIT: The header is the first item in the list returned by
`self.contents()`.
        # Return the first element from that list. The first element of a list is
at index 0.
```

**Sample output of Gemini 2.5 Pro solving the original problem**

*(Other parts of the code omitted for brevity)*

```python
    def header(self) -> list[str]:
        return self.csv.split("\n")[0].split(",")
```

**Sample output of Phi-4 solving the original problem**

```python
    def header(self) -> list[str]:
        """Returns the first row of a CSV file as a list of strings."""
        if not self.csv.strip():
            return []  # Return an empty list if csv is empty or only contains
whitespace
        first_line = self.csv.split("\n")[0]
        return first_line.split(",")
```

**Sample output of Phi-4 following the annotations**

```python
    def header(self) -> list[str]:
        # Get all the rows as a list of lists
        all_rows = self.contents()

        # Return the first row (header)
        return all_rows[0]
```

**Sample output of starcoder2:instruct solving the original problem**

```python
    def header(self) -> list[str]:
        """Returns the first row of a CSV file as a list of strings."""
        return self.contents()[0]
```

**Sample output of starcoder2:instruct following the annotations**

```python
    def header(self) -> list[str]:
        rows = self.contents()
        return rows[0]
```

**Sample output of Comma solving the original problem**

```python
    def header(self) -> list[str]:
        return self.csv.split(",")[0]
```

This output is incorrect as it would only return the first field as string.

**Sample output of Comma following the annotations**

```python
    def header(self) -> list[str]:
        return self.contents()[0]
```

---

## B.2 Test case 28 — Password Strength Checker

**Prompt Type** Lazy

**Instructions** Add a function `containsSpecialChar` that checks if a string contains a special character. Update `isPasswordStrong` to check for the presence of a special character in the password.

**Sample output of Gemini 2.5 Pro as annotator**

```
# EDIT: Create a new function here called `containsSpecialChar`. It should accept
one argument, `password`.
# EDIT: Inside the `containsSpecialChar` function, you will check if the password
string contains any special characters.
# 1. First, create a string variable named `special_characters` and assign it a
string of characters to consider special, for example: "!@#$%^&*()-_=+".
# 2. Then, iterate through each character of the `password` string using a for
loop.
# 3. In each iteration, check if the current character is present in your
`special_characters` string.
# 4. If you find a special character, the function should immediately return
`True`.
# 5. If the loop finishes without finding any special characters, it means there
```

```
are none, so the function should return `False` after the loop.

def minLength(password):
    assert type(password) == str
    return len(password) >= 8

def isPasswordStrong(password):
    # EDIT: Update this return statement. A strong password must now also contain a
special character.
    # EDIT: The function should return `True` only if `minLength(password)` is true
AND the new `containsSpecialChar(password)` function also returns true.
    # EDIT: You can achieve this by calling both functions and combining their
boolean results with the `and` logical operator.
    return minLength(password)
```

**Sample output of Gemini 2.5 Pro solving the original problem**

*(`minLength` is omitted for brevity unless modified)*

```
def containsSpecialChar(password):
    assert type(password) == str
    special_chars = "!@#$%^&*()-+?_=,<>/."
    return any(c in special_chars for c in password)

def isPasswordStrong(password):
    return minLength(password) and containsSpecialChar(password)
```

**Sample output of Phi-4 solving the original problem**

```
import string

def minLength(password):
    """Checks if the password has at least 8 characters."""
    assert type(password) == str
    return len(password) >= 8

def containsSpecialChar(password):
    """
    Checks if the password contains at least one special character.

    Special characters are defined as any character in string.punctuation.
    """
    assert type(password) == str
    return any(char in string.punctuation for char in password)

def isPasswordStrong(password):
    """Determines if a password is strong by checking its length and presence of
special characters."""
    return minLength(password) and containsSpecialChar(password)
```

**Sample output of Phi-4 following the annotations**

```
def containsSpecialChar(password):
    special_characters = "!@#$%^&*()-_=+"
```

```
    for char in password:
        if char in special_characters:
            return True

    return False

def isPasswordStrong(password):
    return minLength(password) and containsSpecialChar(password)
```

**Sample output of starcoder2:instruct solving the original problem**

```
def isPasswordStrong(password):
    if containsSpecialChar(password):
        return True
    else:
        return minLength(password)

def containsSpecialChar(string):
    special_chars = ['!', '@', '#', '$', '%', '^', '&', '*']

    for char in string:
        if char in special_chars:
            return True

    return False
```

**Sample output of starcoder2:instruct following the annotations**

```
def containsSpecialChar(password):
    special_characters = "!@#$%^&*()-_=+"

    for char in password:
        if char in special_characters:
            return True

    return False
```
*(Entire output – original code were removed)*

**Sample output of Comma solving the original problem**

```
def containsSpecialChar(password):
    return any(char.islower() for char in password)

def isPasswordStrong(password):
    return minLength(password) and containsSpecialChar(password)
```

`containsSpecialChar` here checks whether the password contains any lowercase character – and not whether the password has special characters within.

**Sample output of Comma following the annotations**

*(Comma did not remove the `EDIT` comments - it is removed here for brevity. However, it did remove the `minLength` function rendering this code non-functional)*

```python
def containsSpecialChar(password):
    assert type(password) == str
    special_characters = "!@#$%^&*()-_=+"
    for char in password:
        if char in special_characters:
            return True
    return False

def isPasswordStrong(password):
    return minLength(password) and containsSpecialChar(password)
```